\documentclass[a4paper,twocolumn,english,aps, superscriptaddress, preprintnumbers, showpacs, nofootinbib]{revtex4}
\usepackage[T1]{fontenc}
\usepackage[latin1]{inputenc}
\usepackage{amsmath}
\usepackage{amssymb}

\makeatletter
\usepackage{amsfonts}
\usepackage{latexsym}
\usepackage[T1]{fontenc}
\usepackage{ae,aecompl}

\usepackage{babel}
\makeatother
\begin{document}

\preprint{EPFL-ITS-16.2004; UCL-IPT-04-11}

\title{Primordial constraint on the spatial dependence of the Newton constant}

\author{V. Boucher}

\author{J.-M. Gérard}

\affiliation{Institut de Physique Théorique, Université catholique de Louvain,
B-1348 Louvain-la-Neuve, Belgium}

\author{P. Vandergheynst}

\author{Y. Wiaux}

\email{yves.wiaux@epfl.ch}

\affiliation{Signal Processing Institute, Swiss Federal Institute of Technology,
CH-1015 Lausanne, Switzerland}

\date{July 2004}

\begin{abstract}
A Nordtvedt effect at cosmological scales affects the acoustic oscillations
imprinted in the cosmic microwave background. The \emph{gravitational
baryonic mass density} of the universe is inferred at the first peak
scale from WMAP data. The independent determination of the \emph{inertial
baryonic mass density} through the measurement of the deuterium abundance
in the framework of standard big bang nucleosynthesis leads to a new
constraint on a possible violation of the strong equivalence principle
at the recombination time.
\end{abstract}


\maketitle
The cosmic microwave background (CMB) anisotropies provide a unique
laboratory for achieving precision cosmology. The recent analyses
of the corresponding temperature (and polarization) power spectrum,
combined with other cosmological tests, lead to a coherent picture
of the structure, energy content, and evolution of our universe. The
corresponding cosmological parameters are already determined with
a rather high precision by the one-year WMAP data \cite{page,spergel}.
However, in this context full credit may not be given to the concordance
cosmological model before the theoretical hypotheses on which it is
based are tested, notably through a thorough analysis of the CMB.
The inflation scenario \cite{bouchet,coles} and the cosmological
principle must be questioned \cite{coles,hansen,hajian}, as well
as, perhaps most fundamentally, the theory of gravitation itself on
which cosmology is developed, \emph{}namely general relativity. In
this letter we investigate the effect of a strong equivalence principle
violation induced by the spatial variation of the newtonian gravitational
coupling at cosmological scales. The corresponding cosmological Nordtvedt
effect on the CMB provides a new test of general relativity, at the
recombination time.

The equivalence principle, postulating the universality of free fall,
is an important fundament of any theory of gravitation. It is however
implemented at different levels in different theories, thus distinguishing
them from one another in their most fundamental structure. This distinction
may be structured in terms of the already deeply discussed question
of the spacetime variation of fundamental coupling constants \cite{uzan},
such as the fine structure constant $\alpha$ for electromagnetic
interactions (but also the speed of light $c$, the weak and strong
interaction couplings, etc.) or the newtonian gravitational constant
$G$. This spacetime variability of coupling constants is natural
in the framework of the present unified theories for the fundamental
interactions, such as string theories. In this theoretical framework,
auxiliary fields of gravitation indeed appear beyond the tensor metric
field, upon which the fundamental coupling constants naturally depend.
The variation of $\alpha$ in gravitational fields violates the universality
of non-gravitational experiments in free fall, the so-called Einstein
equivalence principle. Many constraints have been established on the
variation of $\alpha$ at low redshifts ($z\leq4$), at the big bang
nucleosynthesis (BBN) epoch, and lately at recombination time through
the analysis of its influence on the CMB anisotropy spectra \cite{rocha}.
Recent evidence for a time variability of the fine structure constant
has been found through the analysis of quasar absorption line spectra,
while the same data indicate no spatial variation \cite{murphy}.
Any spacetime variation of $G$ violates the universality of gravitational
experiments in free fall, known as the strong equivalence principle
(SEP). This principle actually distinguishes general relativity from
any other theory of gravitation of interest, such as scalar-tensor
or vector-tensor alternatives \cite{gerard,damour92}. The time variation
of $G$ originally proposed by Dirac has been extensively analyzed,
leading to constraints at present times, at the BBN epoch, and lately
as well at recombination time through the analysis of the CMB \cite{will,copi,kneller,catena,nagata,chen}.
The possible scale dependence of $G$ has been envisaged \cite{dvali04,dvali01,bertolami99,bertolami96}.
Recent studies also contemplate a dependence of the strength of the
gravitational coupling on the nature of interacting particles \cite{masso,barrow}.
Here, we consider the explicit spatial dependence of the Newton constant,
never yet studied at cosmological scales.

If the newtonian gravitational coupling is a function of the position
$x$ in spacetime, $G\rightarrow G(x)$, the mass $m$ of a compact
body also depends on the position through its internal gravitational
binding energy. An effective action for the geodesic motion of compact
bodies may therefore be defined as $S_{mat}=-c\int m(x)ds$. Energy-momentum
conservation is therefore broken through the introduction of a source
term in the general covariant conservation equations. We adopt the
corresponding covariant expression as our mathematical implementation
of a possible SEP violation:\begin{eqnarray}
T_{\quad\shortmid\nu}^{\mu\nu} & = & G^{,\mu}\frac{\partial T}{\partial G}\quad,\label{eq:1}\end{eqnarray}
where $T$ is the trace of the energy-momentum tensor $T^{\mu\nu}$.
This relation defines a simple modification of the theory of gravitation,
which singles out the spacetime dependence of the Newton constant
as the unique perturbation to the gravitational interaction defined
in general relativity. In the following, we restrict ourselves to
a pure spatial dependence of the newtonian coupling.

The dependence of the gravitational coupling on the spatial position
$\vec{x}$ is parametrized through the relation $G(\vec{x})=G_{0}(1+\eta_{g}V(\vec{x})/c²)$,
where $V(\vec{x})$ stands for the gravitational potential at the
point considered, $G_{0}$ is the background value of the gravitational
constant in the absence of this potential, and $\eta_{g}$ is the
amplitude of the SEP violation. Let us define the compactness $s$
of a body as the sensitivity of its mass relative to $G$. It is equivalently
given by the ratio of its internal gravitational binding energy $E_{g}$
to its total mass energy: $s=-d\ln m/d\ln G=|E_{g}|/mc^{2}$. From
the definition (\ref{eq:1}), one may easily show that the newtonian
acceleration of a body in a gravitational field now explicitly depends
on its proper sensitivity $s$. In other words, the SEP violation
induces a departure of the gravitational mass $m_{g}$ of a body relative
to its inertial mass $m$, proportionally to its own compactness:
$m_{g}=m(1-\eta_{g}s)$. The SEP violation defined in (\ref{eq:1})
therefore reduces to the well-known Nordtvedt effect \cite{nordtvedt,will},
once we neglect the possible time variation of $G$.

We now have to understand this effect on cosmological grounds and
analyze its particular implication for the CMB physics. In the primordial
universe the photon gas may be considered to be tightly coupled to
baryons through the interplay of Compton scattering and Coulomb interaction.
We may therefore consider a photon-baryon plasma in the gravitational
potentials produced by the dominant cold dark matter component of
the expanding universe. The cosmic microwave background radiation
observed today corresponds to a snapshot of the photon gas decoupled
from the rest of the universe at the time of the last scattering.
The structure of the anisotropy distribution on the sky today is defined
by the multiple physical phenomena which governed the evolution of
the plasma before recombination. The well known acoustic peaks in
the corresponding temperature power spectrum originate from electromagnetic
acoustic oscillations of the photon gas. Odd and even peaks respectively
correspond to scales which had reached maximum compression and rarefaction
at the time of last scattering in potential wells (conversely in potential
hills). However, the action of gravity is also introduced through
a purely newtonian coupling of the baryonic content of the plasma
to the dark matter potentials. The effect of this coupling is to shift
the equilibrium point of the oscillations toward more compressed states
in potential wells (rarefied states in potential hills). Consequently,
the height of odd peaks relative to even peaks is enhanced proportionally
to the total baryon weight in the dark matter potentials \cite{hu01,hu96,hu95a,hu95b}.
The temperature power spectrum peaks height therefore bears the imprint
of a possible SEP violation through the Nordtvedt effect as it essentially
originates from a gravitational interaction and thus depends on a
gravitational, rather than inertial, baryonic mass density:\begin{eqnarray}
R_{g}\left(s_{b},\eta_{g}\right) & = & R\left(1-\eta_{g}s_{b}\right)\quad.\label{eq:2}\end{eqnarray}
The compactness $s_{b}$ is now associated with a baryon-region seen
as a homogeneous (under the hypothesis of the cosmological principle)
compact body at the relevant cosmological scale. The canonical variable
$R=3\rho_{b}/4\rho_{\gamma}$ stands for the baryonic density $\rho_{b}$
normalized by the photon density $\rho_{\gamma}$, as it still appears
in the continuity and Euler equations derived from (\ref{eq:1}) for
the evolution of the photon-baryon plasma \cite{boucher}. For simplicity,
the SEP violation parameter is assumed to be constant throughout the
cosmological evolution before recombination: $\eta_{g}=\eta_{g}^{*}$,
where the superscript $^{*}$ evaluates quantities at the recombination
time. This approximation is natural in the framework of string-inspired
theories.

The compactness of a homogeneous spherical baryon-region of radius
$L$ and total mass $M_{b}$, calculated as the ratio of the internal
gravitational binding energy over the total mass energy reads: $s_{b}=3GM_{b}/5Lc^{2}=4\pi G\rho_{b}L^{2}/5c^{2}$.
At each instant in the course of the universe expansion, the maximal
size of the radius $L$ is set by the event horizon. This hypothesis
is natural as the event horizon defines at each moment the maximal
distance through which particles may have interacted gravitationally
since the primordial ages of the universe (after inflation), and therefore
the maximal size of a cosmological body. For the sake of the analogy
with the Nordtvedt effect on compact bodies in a gravitational field,
we consider in the following a constant compactness over the course
of the universe evolution until recombination: $s_{b}=s_{b}^{*}$.
Assuming that the time dependent Friedmann-Lemaître equations remain
essentially unchanged, one may justify this hypothesis, knowing that
recombination takes place inside the matter era. In terms of physical
quantities (the Hubble constant, the age of the universe and the relative
baryon density), we then get for the maximal radius\begin{eqnarray}
s_{b}^{1*} & = & \frac{27}{10}\left(H^{0}t^{0}\right)^{2}\Omega_{b}\simeq0.1\quad,\label{eq:3}\end{eqnarray}
where the superscript $^{0}$ evaluates quantities at the present
time. The low baryon density is indeed largely compensated by the
cosmological scales involved to give a non-negligible compactness.
This compactness is the sensitivity to be considered at the scale
of the wavelength $\lambda_{1}$ associated with the first acoustic
peak. The sensitivity of the baryonic body relevant for the subsequent
acoustic peaks ($\lambda_{n}$) scales like $n^{-2}$: $s_{b}^{n*}\simeq0.1n^{-2}$. 

The independent measurements of both the gravitational baryonic mass
density of the universe and its inertial counterpart lead to a constraint
on the SEP in terms of the cosmological Nordtvedt effect defined in
(\ref{eq:2}). On the one hand, the value for the parameter $\Omega_{b}h^{2}$
obtained from CMB data, is understood in first approximation as a
measurement of the relative height of the temperature power spectrum
odd and even peaks. The small contributions of the inertial baryonic
content to the power spectrum, notably through the sound speed in
the primordial plasma itself affecting the peaks position, are neglected
in this approximation. The one-year WMAPext results (i.e. WMAP extended
to the CBI and ACBAR experiments) give $\Omega_{b}h^{2}=(22\pm1)\times10^{-3}$,
essentially measuring the relative height of the first two peaks \cite{page,spergel}.
The corresponding value for the \emph{gravitational} \emph{baryonic
mass density} $R_{g}^{*}(s_{b}^{1*},\eta_{g}^{*})$ of the universe
at last scattering, and at a scale corresponding to the maximum oscillation
wavelength therefore reads: $R_{g}^{*}=0.613\pm0.028$. On the other
hand, from the determination of light element ($D$, $^{3}He$, $^{4}He$,
$^{7}Li$) abundances, the standard BBN theory may infer the \emph{inertial}
\emph{baryonic mass density} of the universe, essentially counting
nuclei on astrophysical scales and through non-gravitational interactions.
The deuterium abundance is extremely sensitive to the primordial baryon
content. Moreover it may only have been produced in significant quantities
during BBN. Its measurement in quasar absorption line systems is therefore
an extremely good probe of the baryon content of our universe. The
most recent estimate of the primordial deuterium-to-hydrogen abundance
ratio reads: $D/H=2.78_{-0.38}^{+0.44}\times10^{-5}$ \cite{kirkman}.
The corresponding baryon content is given through standard BBN by
$\Omega_{b}h^{2}=(21.4\pm2)\times10^{-3}$, or $R^{*}=0.596\pm0.056$.
Combined with the one-year WMAPext value, this measure gives the first
constraint on a possible violation of the SEP through a cosmological
Nordtvedt effect:\begin{eqnarray}
\eta_{g}^{*} & = & -0.3\pm1\quad.\label{eq:4}\end{eqnarray}
Large systematic uncertainties still affect the other light element
abundance estimation, therefore leading to less reliable assessments
\cite{fields}. These measurements, taken at face value only affected
by statistical errors, would indicate a sizeable SEP violation.

To be more accurate, the constraint (\ref{eq:4}) should be determined
through a best fit of our modified theory (\ref{eq:1}) and present
experimental data, taking into account the substitution (\ref{eq:2})
in the plasma evolution equations before recombination. Also notice
that, in the more complete approach of a specific scalar-tensor or
vector-tensor alternative to general relativity, our cosmological
Nordtvedt effect would no longer remain the only new effect. The introduction
of auxiliary gravitational fields indeed affects the fundamental nature
of gravitation and notably leaves complex signatures in the CMB \cite{catena,nagata,chen}
as well as in the BBN \cite{catena,carroll,damour99}. This would
inevitably modify the proposed constraint. In such a framework, the
bound on $\eta_{g}^{*}$ could also be run backward or forward over
cosmological timescales in terms of the evolution of the auxiliary
fields themselves. This would allow its comparison, either with theoretical
predictions on initial conditions (string theories suggest a violation
parameter of order unity at the outset of the radiation era), or with
present experimental constraints ($\eta_{g}^{0}\leq1\times10^{-3}$
\cite{will}). Finally, other implications of a cosmological Nordtvedt
effect should also be studied beyond its impact on the CMB, at different
epochs of the universe evolution. 

The present considerations are further developed on the ground of
theory and data analysis in \cite{boucher}.

The authors wish to thank P. J. E. Peebles and N. Sugiyama for interesting
comments and discussions. The work of V. B. and J.-M. G. was supported
by the Belgian Science Policy through the Interuniversity Attraction
Pole P5/27. Y. W. also acknowledges support of the european Harmonic
Analysis and Statistics for Signal and Image Processing research network.

\end{document}